\documentstyle[aps,pra,twocolumn,floats,epsfig]{revtex}
\begin{document}
\wideabs{
\title{Observation of transient gain without population inversion in a laser-cooled
rubidium lambda system}
\author{S. R. de Echaniz,$^{1}$ Andrew D. Greentree,$^{1}$ A. V. Durrant,$^{1}$ D.
M. Segal,$^{2}$ J. P. Marangos,$^{2}$ and J. A. Vaccaro$^{3}$}
\address{$^{1}$Quantum Processes Group, Department of Physics and Astronomy, The Open
University,\\
Walton Hall, Milton Keynes, MK7 6AA, United Kingdom.\\
$^{2}$Laser Optics and Spectroscopy Group, Blackett Laboratory, Imperial
College of Science, Technology and Medicine,\\
Prince Consort Road, London SW7 2BW, United Kingdom.\\
$^{3}$Department of Physics and Astronomy, University of Hertfordshire,\\
College Lane, Hatfield, AL10 9AB, United Kingdom.}
\date{\today}
\maketitle

\begin{abstract}
We have observed clear Rabi oscillations of a weak probe in a strongly
driven three-level lambda system in laser-cooled rubidium for the first
time. When the coupling field is non-adiabatically switched on using a
Pockels cell, transient probe gain without population inversion is obtained
in the presence of uncoupled absorptions. Our results are supported by
three-state computations.
\end{abstract}

}

In the last decade there has been growing interest in the study of atomic
coherence effects in multilevel atoms, especially coherent population
trapping \cite{bib:CPT}, electromagnetically induced transparency (EIT) \cite
{bib:EIT}, slow light propagation \cite{bib:UltraSlow}, and optical gain
without inversion. Reviews of lasing without inversion have been given by
Kocharovska and Scully \cite{bib:GWI} and more recently by Mompart and Corbal%
\'{a}n \cite{bib:LWI}. Most of this work has been in the steady state regime
and the fast picosecond pulsed regime \cite{bib:PicoSec}, but there have
been several studies of transient effects where the coherence is switched on
or off rapidly in a time that is short compared to decay rates and Rabi
frequencies, i.e., in the non-adiabatic regime. An early experiment by Fry 
{\it et al.} \cite{bib:LWIvPT} demonstrated several transient effects in a
three level lambda system realized in a sodium vapor cell. Their
observations included transient gain of a strong field when a weak field was
switched on by a Pockels cell in the presence of incoherent pumping to the
upper level, but without population inversion. Subsequently, several
theoretical studies of nondegenerate three-level lambda, V, and cascade
systems have predicted transient ringing of a probe beam with gain when a
strong coupling beam is switched on \cite{bib:TransLambda,bib:TransEITTh}.
This predicted transient gain can occur without population inversion on the
probe transition, and occurs with and without incoherent pumping. This
ringing has also been explained by Vaccaro {\it et al.} \cite{bib:StochWave}
via stochastic wavefunction diagrams. Transient ringing of a three level
system with gain has been observed in the radio-frequency regime in
nitrogen-vacancy centres in diamond samples \cite{bib:DressStateNut} but not
yet in the optical region. Since the ringing occurs at approximately the
Rabi frequency of the strong field, experiments in the optical region would
need to be carried out in a Doppler-free configuration or ideally in a
laser-cooled sample to avoid the Doppler effects masking the coherent
effects.

In our previous work \cite{bib:TransEITExp} we observed the transient
approach to EIT in a cold rubidium lambda system in a magneto-optical trap
(MOT). When the coupling field was switched on by a Pockels cell, the rapid
rise in probe transparency exhibited an overshoot before settling down to
the steady state. This overshoot was interpreted as the first Rabi half
cycle in the transient ringing, but it did not reach gain owing to
absorption on uncoupled Zeeman transitions and two-photon dephasing effects.
In the experiments reported in this paper the coupling field intensity has
been increased to the point where a clear Rabi ringing cycle reaching well
into gain is observed. The results are supported by density matrix
computations of a three level lambda system supplemented by observational
data on the strengths of uncoupled absorptions and the effects of the MOT
fields.

\begin{figure}[tb]
\epsfig{file=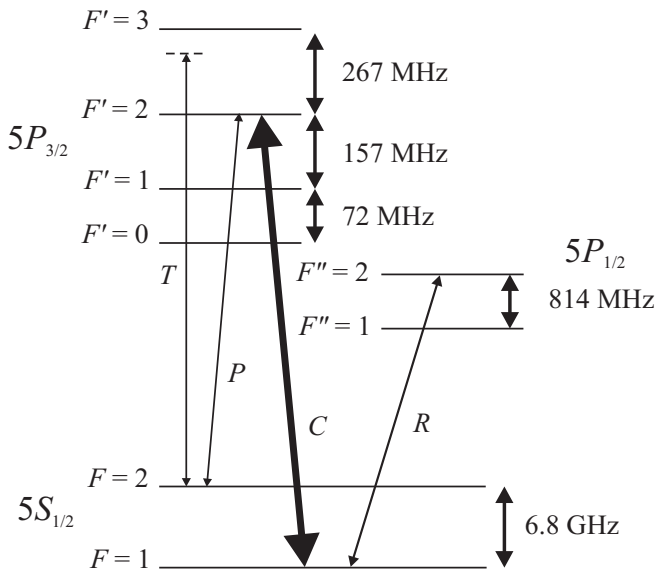,width=86mm}
\\
\caption{The $^{87}$Rb system in which the experiments are carried out. The
atoms are trapped and cooled in a magneto-optical trap by trapping fields $T$ and
$R$. The trapping beams are detuned by $-13\mathop{\rm MHz}$ from the $F=2$
to $F^{\prime}=3$ transition of the D$_{2}$ line. The probe beam $P$ and the
coupling beam $C$ form a $\Lambda$ EIT system.}
\label{fig:RbLevels}
\end{figure}

The lambda system we have studied is formed by the weak probe field $P$ and
the strong coupling field $C$ shown in Fig.\ \ref{fig:RbLevels}. The $^{87}$%
Rb sample was cooled in a standard MOT with trapping field $T$ and repumping
field $R$, similar to the one used in our previous work on EIT \cite
{bib:EITExp}. All laser fields were derived from external-cavity
grating-controlled $780%
%TCIMACRO{\unit{nm}}%
%BeginExpansion
\mathop{\rm nm}%
%EndExpansion
$ laser diodes in master and master-slave arrangements, acting on the
hyperfine transitions of the $5S_{1/2}$ to $5P_{3/2}$ $D_{2}$ line, with the
exception of the trap's repumping beam $R$, which is locked to the $5S_{1/2}$
to $5P_{1/2}$ $D_{1}$ line at $795%
%TCIMACRO{\unit{nm}}%
%BeginExpansion
\mathop{\rm nm}%
%EndExpansion
$. The frequency of each master laser is monitored by saturated absorption
in a Rb cell at room temperature and can be locked via electronic feedback.
In all master-slave arrangements, there is an acousto-optic modulator which
shifts the frequency of the slave relative to the frequency of the master.

The trapping lasers $T$ are derived from a master-slave system. They are
locked and detuned by $-13%
%TCIMACRO{\unit{MHz}}%
%BeginExpansion
\mathop{\rm MHz}%
%EndExpansion
$ from the $F=2$ to $F^{\prime }=3$ transition. Their diameter is $\approx 1%
%TCIMACRO{\unit{cm}}%
%BeginExpansion
\mathop{\rm cm}%
%EndExpansion
$ and the total average intensity in the cold sample is $\approx 50%
%TCIMACRO{\unit{mW}}%
%BeginExpansion
\mathop{\rm mW}%
%EndExpansion
/%
%TCIMACRO{\unit{cm}}%
%BeginExpansion
\mathop{\rm cm}%
%EndExpansion
^{2}$. The repumping beam $R$ is locked to the $F=1$ to $F^{\prime \prime
}=2 $ transition and has a diameter of $\approx 1%
%TCIMACRO{\unit{cm}}%
%BeginExpansion
\mathop{\rm cm}%
%EndExpansion
$ and an intensity of $\approx 0.2%
%TCIMACRO{\unit{mW}}%
%BeginExpansion
\mathop{\rm mW}%
%EndExpansion
/%
%TCIMACRO{\unit{cm}}%
%BeginExpansion
\mathop{\rm cm}%
%EndExpansion
^{2}$. The probe beam $P$ can be locked to or scanned across the $F=2$ to $%
F^{\prime }=2$ transition by piezo-control of the external cavity. $P$ has
an average intensity $\approx 0.03%
%TCIMACRO{\unit{mW}}%
%BeginExpansion
\mathop{\rm mW}%
%EndExpansion
/%
%TCIMACRO{\unit{cm}}%
%BeginExpansion
\mathop{\rm cm}%
%EndExpansion
^{2}$ in a diameter $\approx 1%
%TCIMACRO{\unit{mm}}%
%BeginExpansion
\mathop{\rm mm}%
%EndExpansion
$. The coupling beam $C$ is locked on resonance with the $F=1$ to $F^{\prime
}=2$ transition with an average intensity of $\approx 100%
%TCIMACRO{\unit{mW}}%
%BeginExpansion
\mathop{\rm mW}%
%EndExpansion
/%
%TCIMACRO{\unit{cm}}%
%BeginExpansion
\mathop{\rm cm}%
%EndExpansion
^{2}$ in a roughly elliptical profile$\ 2%
%TCIMACRO{\unit{mm}}%
%BeginExpansion
\mathop{\rm mm}%
%EndExpansion
\times 4%
%TCIMACRO{\unit{mm}}%
%BeginExpansion
\mathop{\rm mm}%
%EndExpansion
$.

\begin{figure}[tb]
\epsfig{file=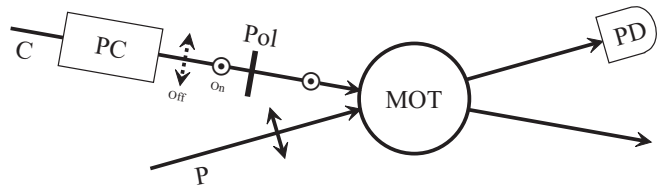,width=86mm}
\\
\caption{Schematic of the experimental arrangement showing the relevant beams
and their polarizations. Field $P$ is linearly polarized in the horizontal plane
whilst field $C$ is linearly polarized in the vertical plane when the
Pockels cell is on, or in the horizontal plane when the Pockels cell is off.
The angle between fields $C$ and $P$ is about $20^{\circ}$. PD is an
avalanche photodiode, PC is a Pockels cell, Pol is a polariser.
The trapping and repumping fields are not shown for clarity.}
\label{fig:ExpSetup}
\end{figure}

The experimental setup is shown schematically in Fig.\ \ref{fig:ExpSetup}.
The probe $P$ is linearly polarized in the horizontal plane whilst the
coupling field $C$ is linearly polarized in the vertical direction. The
probe propagates at an angle of about $20%
%TCIMACRO{\UNICODE[m]{0xb0}}%
%BeginExpansion
{{}^\circ}%
%EndExpansion
$ with respect to $C$, which was found to give a good overlap of the probe
with the coupling field in the cold sample. It can be shown \cite
{bib:V-coupled} that this arrangement of polarizations is equivalent to
three separate sets of Zeeman levels in $\Lambda $ configurations plus
Zeeman transitions of the probe uncoupled by $C$. A Pockels cell and a
polarizer are placed in the path of $C$ so that this beam can be switched on
and off.

\begin{figure*}[p]
\epsfig{file=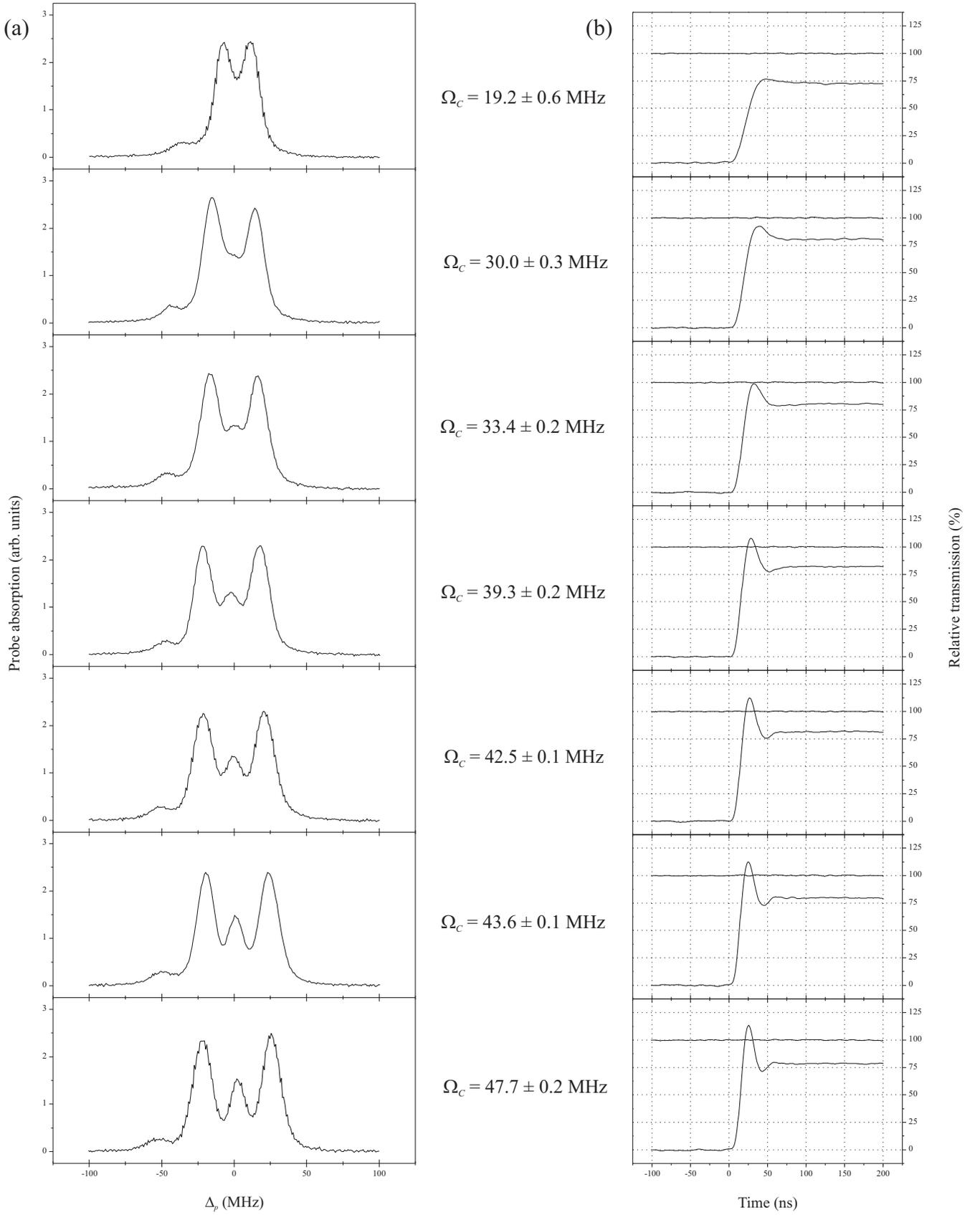,width=7in}
\\
\caption{(a) Steady state probe absorption spectra with $C$ on and locked to the
light-shifted transition of the atoms in the MOT. (b) Probe transient transmission
as $C$ is turned on at time $t=0$. Each trace is an average over $200$ scans.}
\label{fig:Transients}
\end{figure*}

Figure \ref{fig:Transients}(a) shows the steady state probe absorption
versus probe detuning $\Delta _{p}$ for different coupling field Rabi
frequencies as the probe beam is scanned across the $F=2$ to $F^{\prime }=2$
transition. It is seen that the spectrum consists of a central peak situated
between the two Autler-Townes peaks of a standard EIT profile and a small
peak to its red side. The central peak in the spectrum is caused by $P$
probing Zeeman sublevels that are not coupled by $C$. We identified this
peak as the uncoupled absorption peak in our previous work \cite
{bib:V-coupled}. The trapping beams form a V-type EIT system with $P$,
splitting each peak in two, which gives rise to the small red-detuned peak.
We note that, in the present work, it was necessary to shift the frequencies
of $C$ and $P$ to take account of the light shifts caused by the strong
coupling field $C$ and the trapping beams $T$ as described in \cite
{bib:V-coupled}. The shifts of $C$ varied with $\Omega _{C}$ with a maximum
value of $-7%
%TCIMACRO{\unit{MHz}}%
%BeginExpansion
\mathop{\rm MHz}%
%EndExpansion
$, while the shift of the probe was $4%
%TCIMACRO{\unit{MHz}}%
%BeginExpansion
\mathop{\rm MHz}%
%EndExpansion
$. Finally, we note that the linewidths of our spectra are well modelled by
inclusion of the broadening due to beam profile inhomogeneities, variations
of Clebsch-Gordan coefficients between different Zeeman transitions, and the
spread of intensity in the standing wave field of the counterpropagating
trapping beams in the MOT which causes a spread in the detuning of the probe
field due to the spatially varying light shifts.

The transient experiments were performed with the probe field locked to the $%
F=2$ to $F^{\prime }=2$ transition of the cold sample and non-adiabatically
turning on the coupling field in less than $6%
%TCIMACRO{\unit{ns}}%
%BeginExpansion
\mathop{\rm ns}%
%EndExpansion
$ using a Pockels cell driven by a pulse $10%
%TCIMACRO{
%\TeXButton{µs}{\mathop{\rm \mu s}%
%}}%
%BeginExpansion
\mathop{\rm \mu s}%
%
%EndExpansion
$ long with a repetition rate of $10%
%TCIMACRO{\unit{Hz}}%
%BeginExpansion
\mathop{\rm Hz}%
%EndExpansion
$. As the Pockels cell is switched on, it rotates the polarization of the
coupling beam, which then passes through a polarizer that selects only the
rotated polarization (see Fig.\ \ref{fig:ExpSetup}).

The steady-state probe absorption without the coupling field is $\approx
20\% $. Figure \ref{fig:Transients}(b) shows the probe transmission relative
to this initial value, taken to be $0\%$, when $C$ is turned on at time $t=0$
for various coupling field Rabi frequencies $\Omega _{C}$. It is to be noted
that the probe transmission relaxes approximately to the $80\%$ level
instead of the $100\%$ level as predicted by \cite
{bib:TransLambda,bib:TransEITTh}. This is mainly due to the uncoupled
absorption peak and, to a lesser degree, dephasing mechanisms and
inhomogeneities which reduce the visibility of the transparency window, as
seen in the respective steady state EIT traces in Fig.\ \ref{fig:Transients}%
(a). Another important feature of these traces is the fact that only one
whole Rabi cycle is seen, mainly because of the line-broadening caused by
the spread in trapping beam intensities mentioned above. This is in good
agreement with our model when the spread in trapping beam intensity is taken
into account.

\begin{figure}[tb]
\epsfig{file=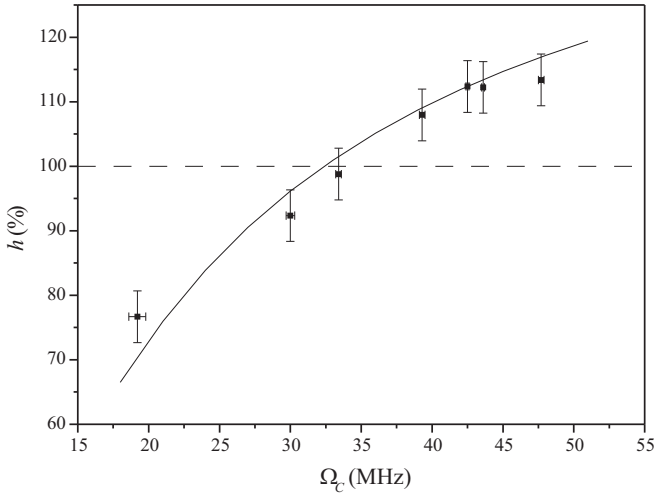,width=86mm}
\\
\caption{The height $h$ of the first Rabi cycle peak plotted against
$\Omega _{C}$ when $C$ is locked to the light-shifted transition of the atoms
in the MOT. The solid line is a theoretical model.}
\label{fig:Heights}
\end{figure}

Despite the above two limiting effects, it is possible to see as much as $%
15\pm 5\%$ gain for maximum $\Omega _{C}$, and an increase in the frequency
of the Rabi cycles with $\Omega _{C}$ as predicted by the theory. Figure \ref
{fig:Heights} shows the different heights $h$ of the first Rabi cycle as a
function of $\Omega _{C}$. Gain is seen in this plot after a threshold Rabi
frequency of $\Omega _{C}\approx 33%
%TCIMACRO{\unit{MHz}}%
%BeginExpansion
\mathop{\rm MHz}%
%EndExpansion
$. Numerically subtracting the effects of the uncoupled absorptions would
yield a maximum gain $\approx 45\%$ and a significantly lower threshold Rabi
frequency. The solid line in Fig.\ \ref{fig:Heights} corresponds to a
theoretical expectation derived from a three-level atom model with an
initial population distribution of $75\%$ in the $F=2$ and $25\%$ in the $%
F=1 $ ground levels, as established from independent absorption
measurements. Our computations show that for these initial conditions there
is no inversion at any time on either the one photon probe transition or the
two photon transition $F=1\rightarrow F^{\prime }=2\rightarrow F=2$.

Theoretical modelling of our system was performed by numerically integrating
the density matrix equations of motion presented earlier \cite
{bib:TransLambda} and incorporating the spread of probe detunings due to the
spatial variation in the light shifts induced by trapping fields. It is
difficult to fully model this spread in trapping field intensity as the
standing wave pattern is critically dependent on the slowly varying, but
unknown, relative phases of the component fields of the trapping beams. For
a fuller description of these effects see \cite{bib:3dStandingWave}.
Experimentally, the mean light shift was corrected by shifting the frequency
of the probe field accordingly, while in the numerical models, a truncated
Lorentzian distribution of detuned probe fields was found to give good
agreement with the experimental results. More accurate modelling of the
actual trap configuration should, however, yield better fits to the data.

We have carried out experiments showing transient ringing of a three-level
lambda system in laser-cooled rubidium when the coupling field is
non-adiabatically turned on. We have observed as much as $15\pm 5\%$
transient gain in the presence of uncoupled absorption. We note that by
working with other configurations, for example, by reversing the roles of $C$
and $P$ in a dark SPOT trap \cite{bib:DarkSPOT}, it should be possible to
eliminate the broadening effects of the trapping fields and the uncoupled
absorptions, thereby observe a larger gain.

We would like to thank the EPSRC for financial support on this project and
Dr. T. B. Smith (Open University) for useful discussions. We would also like
to thank Roger Bence, Fraser Robertson, and Robert Seaton (Open University)
and Shahid Hanif (Imperial College) for technical assistance.


\begin{references}
\bibitem{bib:CPT}  H. Y. Ling, Y-Q. Li, and M. Xiao, Phys. Rev. A {\bf 53},
1014 (1996); E. Arimondo, Prog. Optics {\bf 35}, 257 (1996).

\bibitem{bib:EIT}  S. E. Harris, Phys. Today {\bf 50}, 36 (1997); J. P.
Marangos, J. Mod. Opt. {\bf 45}, 471 (1998); M. Xiao, Y-Q. Li, S-Z. Jin, and
J. Gea-Banacloche, Phys. Rev. Lett. {\bf 74}, 666 (1995).

\bibitem{bib:UltraSlow}  M. M. Kash, V. A. Sautenkov, A. S. Zibrov, L.
Hollberg, G. R. Welch, M. D. Lukin, Y. Rostovtsev, E. S. Fry, and M. O.
Scully, Phys. Rev. Lett. {\bf 82}, 5229 (1999); \ L. V. Hau, S. E. Harris,
Z. Dutton, and C. H. Behroozi, Nature {\bf 397}, 594 (1999).

\bibitem{bib:GWI}  O. Kocharovskaya, Phys. Rep. {\bf 219}, 175 (1992); M. O.
Scully, {\it ibid.} {\bf 219}, 191 (1992).

\bibitem{bib:LWI}  J. Mompart and R. Corbal\'{a}n, J. Opt. B: Quantum
Semicl. Opt. {\bf 2}, R7 (2000).

\bibitem{bib:PicoSec}  A. Nottelmann, C. Peters, and W. Lange, Phys. Rev.
Lett. {\bf 70}, 1783 (1993).

\bibitem{bib:LWIvPT}  E. S. Fry, X. Li, D. Nikonov, G. G. Padmabandu, M. O.
Scully, A. V. Smith, F. K. Tittel, C. Wang, S. R. Wilkinson, and S-Y. Zhu,
Phys. Rev. Lett. {\bf 70}, 3235 (1993).

\bibitem{bib:TransLambda}  Y-Q. Li and M. Xiao, Opt. Lett. {\bf 20}, 1489
(1995); Y. Zhu, Phys. Rev. A {\bf 55}, 4568 (1997).

\bibitem{bib:TransEITTh}  Y. Zhu, Phys. Rev. A {\bf 53}, 2742 (1996); J.
Mompart, C. Peters, and R. Corbal\'{a}n, Quantum Semicl. Otp. {\bf 10}, 355
(1998).

\bibitem{bib:StochWave}  J. A. Vaccaro, A. V. Durrant, D. Richards, S. A.
Hopkins, H. X. Chen, and K. E. Hill, J. Mod. Opt. {\bf 45}, 315 (1998).

\bibitem{bib:DressStateNut}  C. Wei, N. B. Manson, and J. P. D. Martin,
Phys. Rev. Lett. {\bf 74}, 1083 (1995).

\bibitem{bib:TransEITExp}  H. X. Chen, A. V. Durrant, J. P. Marangos, and J.
A. Vaccaro, Phys. Rev. A {\bf 58}, 1545 (1998).

\bibitem{bib:EITExp}  S. A. Hopkins, E. Usadi, H. X. Chen, and A. V.
Durrant, Opt. Comm. {\bf 138}, 185 (1997).

\bibitem{bib:V-coupled}  S. R. de Echaniz, A. D. Greentree, A. V. Durrant,
D. M. Segal, J. P. Marangos, and J. A. Vaccaro, Phys. Rev. A (to be
published); Report No. quant-ph/0102098 (2001).

\bibitem{bib:3dStandingWave}  S. A. Hopkins and A. V. Durrant, Phys. Rev. A 
{\bf 56}, 4012 (1997); S. A. Hopkins, Ph.D. thesis, The Open University,
1995.

\bibitem{bib:DarkSPOT}  W. Ketterle, K. B. Davis, M. A. Joffe, A. Martin,
and D. E. Pritchard, Phys. Rev. Lett. {\bf 70}, 2253 (1993).
\end{references}
\end{document}